\def \yskip{\penalty-50\vskip3pt plus 3pt minus 2pt}
\def \reference{\par \yskip \noindent \hangindent .4in \hangafter 1}
\def \abc#1#2#3#4 {\reference#1, {\sl#2}, {\bf#3}, #4}
\def \blank {\lower 5pt\hbox to 0.75in{\hrulefill}}
\def \kms{~\rm{km}~\rm{s}^{-1}}

\def \cm{~\rm{cm}}
\def \s{~\rm{s}}
\def \km{~\rm{km}}

\def \g{~\rm{g}}
\def \AU{~\rm{AU}}

\def \K{~\rm{K}}

\def \lae{\mathrel{<\kern-1.0em\lower0.9ex\hbox{$\sim$}}}
\def \gae{\mathrel{>\kern-1.0em\lower0.9ex\hbox{$\sim$}}}
\documentstyle[11pt,aasms4]{article}
\begin{document}
%\normalsize
\small

\setcounter{page}{1}
\begin{center} \bf 
DUST FORMATION ABOVE COOL MAGNETIC SPOTS  \\
IN EVOLVED STARS 
\end{center}
%\vspace*{2.0cm}

\begin{center}
Noam Soker \\
Department of Physics, University of Haifa at Oranim\\
%Mathematics-Physics\\
Oranim, Tivon 36006, ISRAEL \\
soker@physics.technion.ac.il \\
and \\
Geoffrey C. Clayton \\
Department of Physics and Astronomy, 
Louisiana State University,\\
Baton Rouge, LA 70803\\
gclayton@fenway.phys.lsu.edu
\end{center}

%\clearpage 
\begin{center}
\bf ABSTRACT
\end{center}

We examine the structure of cool magnetic spots in the photospheres of 
evolved stars, specifically asymptotic giant branch (AGB) stars and 
R Coronae Borealis (RCB) stars. 
We find that the photosphere of a cool magnetic spot will be above 
the surrounding photosphere of AGB stars, opposite to the situation in 
the sun.  
This results from the behavior of the opacity, which increases with 
decreasing temperature, opposite to the behavior of the opacity near 
the effective temperature of the sun. 
We analyze the formation of dust above the cool magnetic spots, 
and suggest that the dust formation is facilitated by strong shocks, 
driven by stellar pulsations, which run through and around the spots.  
The presence of both the magnetic field and cooler temperatures make 
dust formation easier as the shock passes above the spot.  
We review some observations supporting the proposed mechanism, and 
suggest further observations to check the model.


%\clearpage 

% ======================================================================
\section{INTRODUCTION}
% ======================================================================

Several studies in the past suggested that dust can form more easily above 
cool spots in evolved stars.  
Frank (1995) conducted a detailed study of dust formation above cool 
asymptotic giant branch (AGB) starspots, and showed that the mass loss 
rate above the spots increases, though the terminal wind velocity does not 
change much. 
However, Frank does not discuss the source of the cool starspots.  
Schwarzschild (1975) suggested that cool regions in red giants are 
formed by very large convective elements.
Polyakova (1984) suggests two antipodal active magnetic 
regions over 
which dust forms to explain the light and polarization variations in the 
M supergiant $\mu$ Cep.  The spots rotate
with the star and cause the observed light and polarization variations. 
She finds that a rotation period of about 20 years and an activity cycle of about 2.5 years 
fit 
the 
observations.   
Clayton, Whitney, \& Mattei (1993) suggest that the intensive dust 
formation close to the photosphere of R Coronae Borealis (RCB) stars 
can be facilitated by cool magnetic spots.  
In a recent paper, Soker (1998) proposed a scenario in which the 
axisymmetrical mass loss during the high mass loss rate phase at 
the end of the AGB, which is termed the superwind, results from dust 
formation above cool magnetic spots. 
He further argues that this scenario has the advantage that it can operate 
for very slowly rotating AGB and RCB stars, i.e., only $\sim 10^{-4}$ 
times the break up velocity. 
This rotation velocity is $2-3$ orders of magnitude smaller than what 
is required by models were rotation or the magnetic field have a 
dynamical role (Chevalier \& Luo 1994; Dorfi \& H\"ofner 1996;
Ignace, Cassinelli, \& Bjorkman 1996; Garcia-Segura 1997).  
We refer only to elliptical PNs, since bipolar PNs seem to require the 
presence of a close stellar binary companion to their progenitor 
(Soker 1997; Mastrodemos \& Morris 1999).
 
The morphology of planetary nebulae (PNs) and proto-PNs suggest that the transition to the highly 
non-spherical mass loss episode at the end of the AGB is highly nonlinear. 
By nonlinear we mean that a small change in one or more of the properties of the AGB star leads to a 
very large change in the mass loss rate and geometry. 
The mechanism of dust formation via the activity of a dynamo in the envelope of evolved stars is a 
highly nonlinear process (Soker 1998).  This dynamo is not required to form a strong magnetic field.  
A weak magnetic field is enough, as it will be enhanced inside cool spots by a factor of $\sim 10^4$ 
or more, by the convective motion. 
In the sun, for example, the magnetic field in cool spots is 
$\sim 10^3$ stronger than the average magnetic field. 
It cannot reach higher values near the photosphere, since then it greatly exceeds the ambient thermal 
pressure. 
Therefore, it is possible that even if the average magnetic field of the sun were weaker, the intensity 
of the magnetic  field would still reach the same value in cool spots. 
Convective influences on the magnetic field, dust formation and mass loss rate due to dust, are all 
non-linear processes.
For example, the density above the photosphere decreases exponentially with radius (Bedijn 1988; 
Bowen \& Wilson 1991).
Therefore, if the temperature drops a little, dust formation will occur closer to the star where the 
density is much higher (Frank 1995). 
It has been suggested that the superwind results from this increase of the density scale height above 
the photosphere (Bedijn 1988; Bowen \& Wilson 1991).
All the studies described above assume that the photosphere of the cool spot is at the same radius as 
the photosphere of the rest of the star (hereafter the stellar photosphere).
However, we know from the sun that this is not the case.
In the sun, the photospheres of the cool spots are $\sim 2 l_p$ deep in the envelope (Priest 1987, $\S 
1.4.2$D), where $l_p$ is the pressure scale height on the solar photosphere.  This results from the 
lower density and temperature of the spot.
Since the opacity {\it decreases} as temperature decreases, for
conditions appropriate to the solar photosphere, the
photosphere is at higher densities in the spot, which occurs
deeper in the envelope.
As we discuss in $\S 3$ below, the behavior of the opacity is just the opposite in the AGB stars. 
In these stars, the spot will be {\it above} the stellar photosphere.  In $\S 4$ we discuss the formation 
of dust above magnetic cool spots, taking into account the magnetic field, and suggest observations to 
detect cool spots in AGB and RCB stars. 
We summarize in $\S 5$. 
We first turn to examine some observations which support the formation of dust in cool magnetic 
stellar spots ($\S 2$). 
Before doing that we would like to stress that we do not suggest that magnetic activity is the direct 
cause of the enhanced mass loss rate near the equatorial plane. 
We still think that radiation pressure on the dust, the formation of which is facilitated by stellar 
pulsation, does the job.
The magnetic field forms cool spots which further facilitate the formation of dust.
As shown by Soker (1998), the overall magnetic activity is much below the level required by models 
in which the magnetic field has a dynamical role. 

% ======================================================================
% ======================================================================
\section{SUPPORTING OBSERVATIONS}
% ======================================================================
% ======================================================================
% ======================================================================
\subsection {AGB stars}
% ======================================================================
  
Soker (1998) reviews several properties of PNs and AGB stars relevant to the cool magnetic spot 
model. 
The most relevant property that a theory for the formation of elliptical PNs should explain is the 
correlation between the onset of the superwind at the end of the AGB, and the transition to a more 
asymmetrical wind.  In many elliptical PNs, the inner shell, which was formed from the superwind, 
deviates more from sphericity than the outer shell, which was formed from the regular slow wind 
(prior to the onset of the superwind).
In extreme cases, the inner region is elliptical while the outer
shell or halo is spherical (e.g., NGC 6826).
In addition, most ($\sim 75 \%$) of the 18 spherical PNs 
(listed in Soker 1997 table 2) do not have superwind, 
but just an extended spherical halo.  
The correlation between the onset of the superwind and the
onset of a more asymmetrical wind is not perfect, and in some cases both the inner and outer regions 
have a similar degree of asymmetry (e.g., NGC 7662). 
Soker (1998) suggests that magnetic activity may explain this correlation by becoming more 
pronounced at the end of the AGB phase, due to the decrease in the envelope density in the 
convective region (because of mass loss).
Another supporting argument brought by Soker (1998) is the presence of magnetic fields in the 
atmospheres of some AGB stars. 
This is inferred from the detection of X-ray emission from a few M giants
(H\"unsch {\it et al.} 1998). 
Kemball \& Diamond (1997) find a magnetic field at the locations of 
SiO maser emission, which form a ring around TX Cam at a radius of 
$4.8 \AU \simeq 2 R$, and mention the possibility that the 
mass loss occurs in a preferred plane.
They also suggest that ``The fine-scale features [of the Maser image]
are consistent with local outflows, flares or prominences, perhaps
coincident with regions in which localized mass loss has taken place.''
  
We now present more supporting and motivating observations to those 
presented by Soker (1998), through a more careful examination of the 
stellar magnetic activity.
(1)  From the sun we know that during most of the solar cycle, the 
cool spots are concentrated between the equator and latitudes 
$\pm 35^\circ$ (e.g., Priest 1987; $\S 1.4.2$E). 
The model presented by Soker (1998) predicts therefore, that during most 
of the AGB stellar cycle, a higher mass loss rate will occur close to the 
equatorial plane. 
However, at the beginning of a new solar cycle, every $\sim 11$ years,
the cool spots are concentrated at two annular regions 
around latitudes $\sim \pm 30 ^\circ$. 
\newline
(2)     In the sun there are at most several large spots at any given time. 
This means, for dust formation in AGB stars, that the mass loss will be 
enhanced in specific directions, leading to the formation of dense clumps 
in the descendant PN (if spots survive for a long time). 
\newline
(3)     Another property of a stellar magnetic field is that the 
magnetic axis direction can change.  
If the magnetic axis and rotation axis are not aligned, then the 
magnetic axis direction will change during the stellar rotation.  
Another possibility is that the magnetic axis will change in a sporadic way, 
as occurred several times for the Earth's magnetic field.  
  
There is no basic dynamo model to predict the length of
the stellar cycle in AGB stars, the latitude at which spots
appear at the beginning of such a cycle, and the change in the direction 
of the magnetic axis. 
In any case, some morphological features in PNs are consistent with 
an enhanced mass loss rate in two annuli above and below the equator, and 
with a sporadic mass loss rate. 
Let us consider a few examples. 
Some PNs have two annuli, one at each side of the equatorial 
plane, with somewhat higher density than their surroundings.
The PN  K 3-26 (PNG 035.7-05.0; Manchado {\it et al.} 1996), have such
``rings'', with a high
density between the rings, but somewhat lower than in the rings themselves. 
The density in the polar directions is very low. 
Such a structure could be formed from magnetic activity in two 
annuli on the surface of the progenitor AGB star.
Another, more popular explanation, is that a slow wind with a mass 
loss rate which increases toward the equatorial plane, was shaped 
by a fast wind blown by the central star of the PN. 
In the later case, the magnetic activity (to form cool spots) is
only required to enhance the mass loss toward the equator. 
Active annuli may also form two dense rings, which might appear in 
projection as radial condensations in symmetrical configuration, as in 
NGC 6894 (PNG 069.4-02.6; Manchado {\it et al.} 1996; Balick 1987).
Some PNs show loops, arcs, and long condensations extending from the 
shell toward the central star. 
Such features are what we expect from enhanced mass loss rate above 
magnetic cool spots. 
Loops might be caused by the change in direction of the magnetic axis 
and from active annuli on the surface of the progenitor.
Examples of PNs which show clear loops are
A 72 (68PNG 059.7-18.7) and 
NGC 7094 (PNG 066.7-28.2) from Manchado {\it et al.} (1996),
and He2-138 (PNG 320.1-09.6), 
M1-26 (PNG 358.9-00.7), and 
He2-131 (315.1-13.0) from  Sahai \& Trauger (1998). 
Sahai \& Trauger (1998) suggest that the change in direction
of the symmetry axis, and the complicated structures in the inner regions 
of many PNs, may result from multiple sub-stellar (mainly planets) 
companions which interact one after another with the AGB star.  
Although a substellar companion may be the source of the angular 
momentum required to operate the dynamo (Soker 1996; 1998), we think 
that the interaction of several large planets with 
different equatorial planes is very unlikely. 
We prefer sporadic behavior of a stellar dynamo to explain these 
structures in elliptical PNs. 
(Well defined jets in bipolar PNs cannot be explained by our model, 
and probably require stellar companions).
Large long-lasting sporadic magnetic spots might form
dense condensations, as in IC 4593 
(PNG 025.3+40.8; Corradi {\it et al.} 1997),
and A 30 (PNG 208.5+33.2; Manchado {\it et al.} 1996; Balick 1987).
A30 is an interesting PN. 
It has a large, almost spherical, halo, with optically bright, 
hydrogen-deficient, blobs in the inner region (Jacoby \& Ford 1983).  
The blobs, which are arranged in a more or less axisymmetrical shape, 
are thought to result from a late helium shell flash.
Soker (1998) suggests the following explanation according to the magnetic 
cool spots model. 
During the formation of the halo dust was forming far from the stellar 
surface.  
If after the helium flash the formation of dust occurred closer to 
the stellar surface, the process became more vulnerable to 
magnetic activity, resulting in the axisymmetrical mass loss. 
 
% ======================================================================
\subsection {RCB stars}
The RCB stars are rare hydrogen-deficient carbon-rich 
supergiants which undergo very spectacular declines in brightness of up to 8 
magnitudes at irregular intervals as dust forms along the line of sight 
(Clayton 1996).  
There are two major evolutionary models for the origin of RCB stars: 
the double degenerate and the final helium shell flash 
(Iben, Tutukov, \& Yungelson 1996).
The former involves the merger of two white dwarfs, and in the latter a 
white dwarf/evolved PN central star is blown up 
to supergiant size by a final helium flash.
In the final flash model, there is a close relationship between RCB stars 
and PN such as A30, discussed above. The connection between RCB stars and 
PN has recently become stronger, since 
the central stars of three old PN's (Sakurai's Object, V605 Aql and FG Sge; 
Duerbeck \& Benneti 1996; Clayton \& De Marco 1997; Gonzalez et al. 1998) 
have had observed outbursts that transformed them from hot evolved central 
stars into cool giants with the spectral properties of an RCB star. 

Wdowiak (1975) first suggested the possibility that dust in RCB stars
forms over large convection cells which are cooler than the surrounding
photosphere.
Clayton et al. (1993) suggested that a magnetic activity cycle similar to the Solar Cycle could fit in well with the observed 
properties of RCB stars.  
It would provide a mechanism for a semi-periodic variation in dust production, could 
cause cool spots 
over which patchy dust clouds might form, and could be related to the chromospheric 
emission seen in these stars.

There is no direct observational evidence for
a magnetic field in any RCB star.
When in decline, RCB stars do exhibit an emission spectrum that is
often referred to as 'chromospheric' although not all the emission
lines typical to a chromosphere are seen.
Lines associated with transition regions, such as
C II $\lambda$1335, C III] $\lambda$1909 and C IV $\lambda$1550 are also seen 
(Clayton 1996, Maldoni et al. 1999).
 These lines indicate temperatures of $\sim 10^5$ K.
 Models of the transition regions in other stars indicate that acoustic
waves alone cannot provide enough energy to account for the radiation
losses and a small magnetic field must be present (Jordan \& Linsky 1987).
No flares have been observed on an RCB star although Y Mus does exhibit
flickering in its lightcurve (Lawson et al. 1990; Lawson \& Cottrell 1997).
 No X-rays have ever been detected.
 Photometric detection of starspots is difficult due to the presence
of pulsations and dust formation events. 
There is no measurement of the rotation period of an RCB star.
The effect of rotation is not measureable in existing high resolution spectroscopic data (e.g. Pollard, Cottrell, \&
Lawson 1994).
Therefore, the rotation period of these stars is one year or longer.
The pulsation periods of RCB stars lie in the range 40-100 days
(Lawson et al. 1990).
 These are confirmed as pulsational variations by radial velocity
measurements (Lawson \& Cottrell 1997).
 Fourier analysis of RCB light curves do show significant low frequency
($\sim$ 200 d) contributions but they are attributed to couplings of
higher frequency terms or the windowing effect of the observing seasons.
 RY Sgr has two periods seen in its lightcurve of 38 and 55 d
(Lawson et al. 1990).
However, only the 38 d period shows up in radial velocity measurements
(Lawson \& Cottrell 1997).

% ======================================================================
\section{MAGNETIC COOL SPOTS}
% ======================================================================
% ======================================================================
\subsection{The position of the spot's photosphere}
% ======================================================================

Let us examine the structure of a vertical magnetic flux tube as is done 
for sunspots, following, e.g., Priest (1987; $\S 8.4$).  
The biggest uncertainty in the model is the temperature of the cool spot, 
which is also the most important factor for dust formation. 
Keeping this in mind, we will make several simplifying assumptions in 
this section. 
Using the definition of the photosphere as the place where
$\kappa \l \rho_p =2/3$, where $\kappa$ is the opacity, $l$ the density 
scale height and $\rho_p$ the density at the photosphere, 
the pressure at the photosphere is given by 
(e.g.,  Kippenhahn \& Weigert 1990, $\S 10.2$)
\begin{equation}
P_p= 
\frac {2}{3} 
\frac {G M}{R^2} 
\frac {1} {\kappa},
\end{equation}
where $M$ is the stellar mass and $R$ the photospheric radius.            
At the level of accuracy of our calculations, we can take the 
pressure and density scale height at the photosphere to be equal (we do 
not consider here the density inversion region below the photosphere of 
AGB stars; Harpaz 1984). 
The density at the photosphere is given by
\begin{equation}
\rho_p = 
\frac {2}{3} 
\frac {G M \mu m_H}{k} 
\frac {1} {R^2 \kappa T_p},
\end{equation}
where $T_p$ is the photospheric temperature, $k$ the Boltzmann constant, and $\mu m_H$ the mean 
mass per particle. 
Let the subscript $i$ denote quantities in the center of the cool spot, 
and the subscript $e$ quantities outside the spot, 
where the magnetic field can be neglected.
Pressure balance between the spot and its surroundings  reads
\begin{equation}
P_i+P_B = P_e,
\end{equation}
where $P_B$ is the magnetic pressure inside the cool spot.
Derivation with respect to the radial coordinate gives
\begin{equation}
\frac{d P_i}{dr}
= \frac{d P_e}{dr}
- \frac{d P_B}{dr}.
\end{equation}
As in the sun, we assume that the magnetic field lines inside 
the  spot are vertical, and only near the 
photosphere are they open tangentially in order to reduce 
the magnetic pressure.
The magnetic pressure deep in the envelope is of the same order as 
the thermal pressure. 
Near the photosphere the magnetic field has to open up in order for 
the magnetic pressure not to 
exceed the surrounding's thermal pressure (e.g., Priest  1987 $\S 8.4$).
We approximate the magnetic pressure gradient as 
\begin{equation}
\frac{d P_B}{dr} = \frac {P_B - \alpha P_B}{d},
\end{equation}
where $P_B$ is the magnetic pressure on the photosphere of the spot, 
and $\alpha P_B$ is the magnetic pressure at a radius equal to 
the surrounding photospheric (hereafter just photospheric) radius.
$d$ is the radial distance between the photosphere and 
the spot's photosphere. 
In the sun, the spot is deep in the envelope.  In this case, 
$d$ is negative and $\alpha <1$.
In AGB stars, we will find below that the spot photosphere is above 
the photosphere, so that $d>0$ 
is positive and $\alpha > 1$. 
Since the vertical magnetic field lines do not exert radial force
(e.g., Priest 1987 $\S 8.4$), the hydrostatic equilibrium within the 
spot does not include the magnetic pressure gradient
\begin{equation}
\frac {1}{\rho_i} \frac{d P_i}{dr} = 
\frac {1}{\rho_e} \frac{d P_e}{dr}= - g,  
\end{equation}
where $g=GM/R^2$ is the gravity in the photosphere.
Substituting  $dP_i /dr$ from equation (4) into equation (6), and using
equation (5) for $dP_B/dr$, give
\begin{equation}
\frac {1}{\rho_i} 
\left( \frac{d P_e}{dr} - 
\frac {P_B - \alpha P_B}{d} \right) = 
\frac {1}{\rho_e} \frac{d P_e}{dr}.  
\end{equation}
In the photosphere $\rho_p \kappa_p l_p = 2/3$, while in the 
spot's photosphere $\kappa_i l_i \rho_i =2/3$.
From the last two equations we find 
\begin{equation}
\rho_i = \rho_p \frac {\kappa_p l_p}{\kappa_i l_i}.
\end{equation}
We now use our approximation that the pressure scale height is equal to 
the density scale height. 
This is not a bad approximation when there is a steep pressure drop and 
a shallow temperature drop 
as in the photosphere. 
Multiplying and dividing the right hand side of equation (6) by $P_e$
and taking quantities at the photosphere,
and the same for the left hand side with $P_i$, we find
$T_i/l_i = T_p/l_p$.
In the last equality we have assumed that the scale height of the 
surrounding atmosphere near the photosphere of the spot is the 
same as that at the photosphere of the envelope $l_p=l_e \equiv -P/(dP/dr)$.  The density at the photosphere is given by  $\rho_p = \rho_e e^{d/l_p}$, where  
$\rho_e$ is the surrounding envelope density at the radius of the 
spot's photosphere. 
Using the above expressions for $l_i$ and $\rho_p$ in equation (8),
we find 
\begin{equation}
\rho_i = \rho_e \left( \frac {\kappa_p T_p}{\kappa_i T_i}
\right) e^{d/l_p}.
\end{equation}
Since we are considering the spot's photosphere, we will use the subscript 
$s$ instead of $i$ from 
now on. 
Dividing equation (7) by $P_e$, using the definition of the scale height
$l \equiv - P/(dP/dr)$, and substituting for $\rho_i$ from equation (9),
gives after rearranging terms 
\begin{equation}
\frac {\kappa_s T_s}{\kappa_p T_p}
\left[ 1 + \frac {l_p}{d}  \frac{P_B}{P_e} 
( 1 - \alpha ) \right] e^{-d/l} = 1 .
\end{equation}

% ======================================================================
\subsection{The Sun}
% ======================================================================
 
Let us examine the validity of the last equation for the sun.
For the solar photosphere $\rho_p \simeq 10^{-7} \g \cm^{-3}$, $T_p=5800 \K$, gravity $g=2.74 
\times 10^4 \cm \s^{-2}$, and $P=1.2 \times 10^5 {\rm dyne} \cm^{-2}$.
From Alexander \& Ferguson  (1994) and the TOPbase data base 
(Cunto {\it et al.} 1993; Seaton {\it et al.} 1994) we find 
$\kappa \simeq 0.25 \cm^2 \g^{-1}$.
The scale height is $l_p = 280 \km$.
For a typical large cool solar spot, we take $T_s \simeq 3,700 \K$ (e.g., Priest 1987 $\S 1.4$).
By using these opacity tables, we find the photospheric density  and opacity to be $\rho_s \simeq  5 
\times 10^{-7} \g \cm^{-3}$,  and $\kappa_s \simeq 0.05$. 
With these values, we solve equation (10). 
When the pressure gradient is neglected, i.e., $\alpha=1$, the solution is 
$d = -2 l_p$, i.e., the spot is $ \sim 2 l_p \simeq 560 \km $ deep in the 
photosphere.
With $\alpha \ll 1 $ and $P_B \simeq P_e$, the solution is $d=-2.4 l_p$, or 
a depth of $\sim 670 \km$. 
These values are within the range of the depth of the spots in the sun, 
$d \sim - 500-700 \km$, as 
inferred from the Wilson effect (e.g., Priest 1987 $\S 1.4$).
This shows that equation (10) is a good approximation, at least for the sun.  
The reason for the spots being deeper in the envelope is that for the 
typical parameters of the solar photosphere, opacity {\it decreases} 
as temperature decreases.

% ======================================================================
\subsection{AGB stars}
% ======================================================================
In AGB stars, the situation is the opposite of that in the sun.
From the data presented by Alexander \& Ferguson (1994), we find 
that the  opacity drops slightly to a minimum as the temperature drops 
to $\sim 2700 \K$ from $\sim 3,000 \K $, but then sharply increases to a 
value $\gtrsim 50$ times higher at a temperature of $T \lesssim 2100 \K$ 
(all at a constant density). 
The higher opacity in the cool spots means that the density will be lower than in the rest of the 
photosphere. 
Lower density means a somewhat lower opacity,  so that the real increase in opacity will be by a 
factor of $\lesssim 50$. 
Let us consider a specific example. 
 From the definition of the pressure scale height, 
$l \equiv - P/(dP/dr) = P/(\rho g)$, 
we find (in the photosphere), 
\begin{equation}
\frac {l_p}{R}  
\simeq 0.05  % 0.04875
\left( \frac {R}{300 R_\odot} \right)
\left( \frac {T_p}{3,000 \K} \right) 
\left( \frac {M}{0.8 M_\odot} \right)^{-1}
\simeq 0.05
\left( \frac {T_p}{3,000 \K} \right)^{-1} 
\left( \frac {M}{0.8 M_\odot} \right)^{-1}
\left( \frac {L}{6,500 L_\odot} \right)^{1/2},
\end{equation}
where the gravity is $g = G M /R^2$.                        % =0.30445 
We took the mean mass per particle to be $\mu m_H = m_H$, higher than 
for a fully ionized plasma since gas in an AGB star 
photosphere is partially recombined, and RCB stars are hydrogen deficient.
The stellar mass is taken for a typical star on the AGB tip, with envelope 
mass of $0.2 M_\odot$ and a core mass of $0.6 M_\odot$.  
We use the scale height, radius, and temperature as in equation (11), 
to find the photospheric opacity and density.
For the photosphere we get $\rho_p \simeq  10^{-9} \g \cm^{-3}$ and 
$\kappa_p \simeq 5 \times 10^{-4} \cm^{2} \g^{-1}$. 
Following the sun, we take the cool spot to be at a temperature of 
$T_s = 2 T_p/3 = 2,000 \K$. 
We find the density and opacity to be 
$\rho_s \simeq 5 \times 10^{-11} \g \cm^{-3}$ and
$\kappa_s \simeq 1.3 \times 10^{-2}  \cm^2 \g^{-1}$, respectively.  Solving equation (10) with 
these values and taking $\alpha=1$, we obtain $d \simeq 2.85 l_p$.
Taking the magnetic pressure gradient into
account, i.e., $\alpha>1$, will reduce $d$.
Here, $P_B < P_e$, since the pressures are evaluated at the location of the spot photosphere, 
$r=R + d$. 
It will be more convenient to write equation (10) in terms of the pressure ratio  at the stellar 
photosphere, $(P_B/P_e)_{\rm phot}$, rather than at the spot photosphere. 
This ratio is 
\begin{equation}
\left( \frac{P_B}{P_e} \right)
=
\left( \frac{P_B}{P_e} \right)_{\rm phot} 
\frac{ e^{d/l_p} }{\alpha}.
\end{equation}
Rearranging terms in equation (10) gives 
\begin{equation}
e^{-d/l_p}
=
\frac {\kappa_p T_p}{\kappa_s T_s}
+
\frac {l_p}{d}  
\left( \frac{P_B}{P_e} \right)_{\rm phot} 
\left( 1- \frac {1}{\alpha} \right).
\end{equation}
Under the condition that $\kappa_s \gg \kappa_p$, the first 
term on the r.h.s may become much smaller than the second term.  For example, for 
$(P_B/P_e)_{\rm phot} = 0.5$ and $\alpha =2$, and with the other parameters as taken above, the 
solution of equation (13) is $d = 1.5 l_p$.
In this case, the second term on the r.h.s. is $0.167$, while 
the first term is $0.0577$. 
We cannot make $\alpha$ much greater, since then $d$ becomes smaller, and the pressure cannot 
drop by a large factor in such a short distance . 
Going back to neglect the pressure gradient, we examine other temperatures.  At $T_s = 1600 \K$, 
the opacity decreases (relative to $2,000 \K$) to $\kappa_s \simeq 5 \times 10^{-3}  \cm^2 \g^{-1}$, 
and from equation (10) (and for $\alpha=1$) $d= 1.7 l_p$, while at 
$T_s = 2500 \K$ we find  $\kappa_s \simeq 2 \times 10^{-3} \cm^2 \g^{-1}$, and from equation 
(10) $d \simeq  1.2 l_p$. 
We conclude that cool magnetic spots on the surfaces of AGB stars are protruding above the 
photosphere by $1.5-3$ scale heights.  Cool spots at $\sim 2500 \K$ will probably have only a small 
influence, while at $T_s \simeq 1600 \K$ dust is already forming. 
The relevant temperature is $\sim 2,000$, where the spots are $d \simeq 1.5 - 3 l_p \sim 0.1-0.15 R$ 
above the photosphere!  In $\S 4$ we will discuss the implications of the protruding cool spots on 
dust formation and observations.

% ======================================================================
\subsection{RCB stars}
% ======================================================================
 
{\bf Cool RCB stars.}
Cool magnetic spots in cool RCB stars,
($T_{\rm eff} \simeq 5,000-7,000 \K$),  
will be deeper that the surrounding photosphere as in the sun.  
Because of the composition of RCB stars, mainly helium,
the opacities are lower than for solar composition.
From equation (1) we see that the pressure will be higher than that of a solar composition star with 
the same radius, luminosity and mass.  
As an example consider an RCB star of surface temperature
$7,000 \K$, radius $70 R_\odot$, hence luminosity of 
$L=1.05 \times 10^4 L_\odot$, and a mass of $0.6 M_\odot$.
From the table of the TOPbase opacity project (Cunto {\it et al.} 1993),
and the scale height $l_p \simeq 0.03 R$, by equation (11; the mean weight 
per particle in RCB stars is larger than that assumed in equation 11,
$> 1 m_H$, but to first order we still use equation 11)
% $ 1.77 \times 10^{11} \cm$
we find the opacity and density on the photosphere for these hydrogen 
deficient stars ($X=0$, $Z=0.02$) to be  
$\kappa_p \simeq 1.3 \times 10^{-3} \cm^2 \g^{-1}$, and 
$\rho_p \simeq 4 \times 10^{-9}$
(see also model atmospheres by Asplund {\it et al.} 1997). 
For a cool spot of $T_s = 2 T_p/3 = 4,700 \K$,
the opacity and density are
$\kappa_s \simeq 6 \times 10^{-4} \cm^2 \g^{-1}$, and 
$\rho_s \simeq 10^{-8}$, respectively.
From equation 10 we find the depth of the cool spot to be
$d \simeq - 1.2 l_p \simeq - 0.04 R$. 
Taking the pressure gradient into account with $P_B=P_e$ and
$\alpha \ll 1$, we find $d \simeq -1.7 l_p \simeq - 0.05 R$. 
This is deeper than in the sun, since in the sun 
$(l_p/R_\odot) = 4 \times 10^{-4}$, while in cool RCB stars
this ratio is two orders of magnitude higher. 
It will be interesting to conduct numerical simulations, similar to
those of Woitke {\it et al.} (1996), but where the shock waves are
traveling inside the ``pipe'' of the deep, $d \simeq - 0.05 R$,
magnetic spots on cool RCB stars.
This, of course, is beyond the scope of the present paper. 
In $\S 4.1$ we suggest that formation of amorphous carbon dust occurs
as the shock breaks out of the pipe on the surface of the star.
\newline
{\bf Hot RCB stars.}
For a temperature of $T_p = 18,000 \K$, luminosity of $L=10^4 L_\odot$, 
hence $R=10 R_\odot$, and a mass of $M=0.6 M_\odot$,
we find from equation (11) $l_p \simeq 0.01 R$.
The photospheric opacity and density are 
$\kappa \simeq 0.4 \cm^2 \g^{-1}$, and 
$\rho_p \simeq 2 \times 10^{-10}$, respectively. 
% $l = 9.9 \times 10^9 \cm $ 
Opacity tables for a hydrogen deficient atmosphere at 
$T \gtrsim 15,000$ show that the opacity depends very weakly 
(relative to the range of cool RCB stars) on temperature. 
In a small range near $\sim 20,000 \K$ the opacity even increases a 
little as temperature decreases.
 Taking the opacity to be constant, and $T_s=(2/3) \times T_p$, as in 
the sun, we find from equation (10) and for a small magnetic pressure 
gradient (in this case $d$ is small, so we can take $\alpha 
\simeq 1$) $d \simeq - 0.4 l_p$. 
We see that the spot is well inside the photosphere, even when the opacity 
is taken to be constant.  
Below $\sim 15,000 \K$, the opacity decreases steeply, and
if the spot temperature is in this range, then the spot will be
$\sim 1- 2 l_p \simeq 0.01 R$ inside the photosphere,
much shallower than in cool RCB stars. 

% ======================================================================
% ======================================================================
\section{IMPLICATIONS}
% ======================================================================
% ======================================================================
% ======================================================================
\subsection{Dust formation}
% ======================================================================

As a parcel of gas in the wind moves away from the cool spots, 
it starts to get more and more radiation from the hotter surface of 
the star surrounding the spot. 
Therefore, even if initially this parcel is much cooler than the rest 
of the gas in the wind, at some distance from the surface it will be 
at only a slightly lower temperature than the surrounding gas. 
In order to stay much cooler until dust forms, Frank (1995) finds that 
the cool spots should be very large; having radius of a 
few$\times 0.1 R$, where $R$ is the stellar radius.  
There is a problem in forming such large magnetic cool spots.  
This is because the strong magnetic field in cool spots is formed by 
concentrating a weak magnetic field. 
Magnetic flux conservation means that the area from which the weak 
magnetic field is concentrated to the spot is much larger than 
the spot's area. 
This cannot be the case if the magnetic spot is as large as required by 
the calculation of Frank. 
The solution, we think, is that the dust forms very close to the 
cool spot, so that even small spots (but not too small) can form dust. 
We should stress again that the formation of dust above cool spots, 
as suggested here and by Soker (1998), is not intended to replace dust 
formation around the star at several stellar radii (as occurs in AGB stars).  
Our idea is that enhanced dust formation above cool spots increases the 
mass loss rate, and makes the overall mass loss geometry less spherical. 
\newline
{\bf AGB stars.}
The temperature amplitude due to the pulsation of Mira variables can be as high as $\sim 15
\%$
(e.g., Hoffmeister, Richter, \& Wenzel 1985).  This means, that a cool spot of temperature $\sim 
2,000 \K$ can cool to $\sim 1700 \K$. 
The high density of the spot photosphere, means that dust can already forms at this, or a slightly 
lower, temperature.  
Therefore, it is quite possible, that when a large and cool magnetic cool spots forms, large quantities 
of dust formed during the minimum temperature of each pulsation cycle. 
\newline
{\bf RCB stars.}
Such low temperatures are not attainable around cool and hot RCB stars 
even on cool spots. 
However, as described below, Woitke et al. (1996) show that for 
$\rho_s \sim 10^{-13}$ to $10^{-16} \g \cm^{-3}$ conditions allow 
for the condensation temperature for carbon to be reached 
as a shock passes through the atmosphere of the star.  
However, for higher densities, the adiabatic cooling is negligible 
during the reexpansion following the shock so the temperature remains 
near the radiative equilibrium temperature. 
Therefore, the higher densities present inside the cool spot do not 
enhance dust formation in the Woitke scenario. 
The cooler temperatures and magnetic field, by enhancing adiabatic
expansion (see below), may aid dust formation close to the spot where  
the densities are lower than inside the spot.
 
To present our proposed scenario for enhanced dust formation,
in RCB and AGB stars, but in particular in hot RCB stars, we must first 
summarize the effects of shocks as calculated and discussed 
by Woitke {\it et al.} (1996).
Woitke {\it et al.} study the effect of shock waves, excited 
by stellar pulsations, on the condensations of dust around cool RCB stars.  
They consider only the spherically symmetric case, with an effective temperature 
of $T_p=7,000 \K$. 
They examine shocks, which begin to develop somewhere below the photosphere, 
as they run out to several stellar radii.
The shock velocities in their calculations were $20 \kms$ and $50 \kms$.  
Somewhere outside the photosphere, at radius of $\sim 2 R$, the density is 
in the right range for the following cycle to occur.  
$(i)$ As the shock passes through the gas, it compresses the gas by a 
factor of $\sim 6-10$, and heats it by a factor of $\sim 3-10$. 
The compression and heating factors depend mainly on the shock 
velocity.  $(ii)$ due to its higher density, the gas cools very 
quickly to its radiative equilibrium temperature. 
This equilibrium is with the radiation from the photosphere.
$(iii)$ The compressed gas reexpands and its density drops by more than 
an order of magnitude. 
This results in a large adiabatic cooling, which may bring the gas to 
below the dust condensation temperature. 
The decrease in density and hence the adiabatic cooling becomes more 
pronounced as the shock velocity increases.  
In Woitke {\it et al.} calculations, a $50 \kms $ shock results in dust 
formation, while for a $20 \kms$ shock, no dust forms. 
Let us examine what happens during this three stage cycle for gas above a 
cool magnetic spot. 
The pressure equilibrium above the photosphere is given by equation (3), 
i.e., the thermal pressure above the spot plus its magnetic pressure 
equals the thermal pressure of the surroundings (where the 
magnetic pressure is very small). 
$(i)$ As a strong shock moving radially outward passes through a region, 
it compresses the gas by a factor $>4$, and heats it. 
The thermal pressure in the calculations of Woitke {\it et al.} (1996) 
increases by a factor of $\sim 10^2$.
Since the magnetic field lines near the center of the spots are radial 
(e.g., Priest 1987), the magnetic pressure does not increase behind the 
shock.
Therefore, the surrounding post shock pressure exceeds that of the region 
above the spots. 
The surrounding pressure compresses the region above the spot in the 
transverse direction, increasing both the thermal and magnetic pressure 
there, but the magnetic field is still smaller than the thermal pressure.  
Therefore, the region above the spot is compressed by a larger factor than 
the surrounding medium. 
This increases the efficiency of the mechanism studied by
Woitke {\it et al.}.
Both regions, above the spot and the surroundings, reach similar thermal 
states since the magnetic pressure is small. 
$(ii)$ Due to the high density, the gas in the two regions 
cools very fast to its radiative equilibrium temperature.
But above the spot the temperature will be lower. 
$(iii)$ The compressed gas reexpands and its density drops by more than 
an order of magnitude. 
Because of the cooling and the reexpansion, mainly in the radial 
direction, the thermal pressure drops and the magnetic pressure above the 
spots becomes an important, or even the dominant pressure. 
The magnetic field pressure results in a transverse expansion of the 
magnetic field lines.  
Since the gas is partially ionized, it will be practically frozen-in 
to the magnetic field lines, and the gas above the  
spots will expand transversely. 
The net result is that during the adiabatic cooling stage, 
the gas above the spot will reexpand by a larger factor, and hence will 
reach lower temperature. 
 To summarize, cool magnetic spots have two factors which ease dust formation. 
First the temperature is lower, and, second, the magnetic field increases the 
reexpansion, and hence the adiabatic cooling, of the region above the spot.  
Both lower the temperature and density of the spot and the gas above it, and
make the dust formation mechanism studied by Woitke {\it et al.} effective 
closer to the stellar surface. 

\subsection{Observations}

Clayton et al. (1997) found that in a deep decline of R CrB,
the position angle of the continuum polarization was almost flat 
from 1 \micron~to 7000 \AA~but 
then changed rapidly, rotating by $\sim 60^\circ$ between 7000 and 4000 \AA. 
This behavior is strikingly similar to that produced in post-AGB stars 
having an obscuring torus and bipolar lobes of dust.  
These new data strengthen the earlier suggestion that there is a preferred 
direction to the dust ejections in R CrB (Clayton et al. 1995).  
Dust ejections seem to occur predominantly along two roughly orthogonal 
directions consistent with a bipolar geometry.
Another example of asymmetrical mass loss from RCB stars
is the apparent bipolar nebulosity observed
around UW Cen (Pollacco et al. 1991). 
 However, Clayton et al. (1999) find that the shape of the 
nebula changes with time due to changes in the illumination from the star.
More observations are planned to detect and map the morphology of 
shells around RCB stars.

Starspots have been detected and mapped on a number of stars using 
techniques which combine photometry and spectroscopy 
(Vogt \& Penrod 1983; Strassmeier 1988 and references therein). 
 The Doppler Imaging technique uses spectra of sufficient resolution to 
resolve individual stellar lines into several velocity bins. 
 Because RCB stars likely rotate so slowly, extremely high spectral resolution 
would be required for Doppler Imaging.  
But accurate long-term photometric observations can be used to test for the presence of spots. 
The problem of confusion with pulsations and dust formation remain.
The predicted RCB starspots will lie below the photosphere of star 
like those on the Sun and should be distinguishable from spots at 
the level of or higher than the stellar photosphere as we predict for
AGB stars.
Due to the Wilson effect, the spots will be vignetted 
when near the stellar limb affecting the photometric behavior of the 
star (Priest 1987).
 
The main problem of observing cool spots on these evolved stars is
that when the spot is large and long-lived, we predict enhanced
dust formation, which complicates the observation.
 In addition, these stars rotate very slowly, so the rotation period
is likely to be longer than the lifetime of a magnetic cool spot.
 This means that photometric variations due to rotation, are very
hard to detect.
 Therefore, the detection of cool spots is very tricky. 
 In RCB stars a careful observation should be made before a deep
decline, looking for photometric characteristics of a large cool region
(of course, we'll know we observed the right time only after the
decline).
  The spot will form on a dynamical time scale, which for RCB stars
is $\sim 1-2$ months. 
 Pulsation, as stated above, will complicate things considerably.
In any case, broadband photometry of an RCB star  obtained
over a few months before a decline should be carefully compared with
observations of the same time span in quiet times. 

 In AGB stars the situation is much more complicated. 
 In addition to dust formation above cool spots, the star forms
large quantities of dust further out due to the large amplitude stellar
pulsations and the cool photosphere. 
 These stars tend to be obscured by dust. 
 We suggest the following type of observations.
 The target stars should be on the upper AGB, preferentially Carbon stars, 
but before dust obscuration.  
 At this stage magnetic activity starts to become significant 
(Harpaz \& Soker 1999), and the cool spots are expected 
live for a few weeks to a few months. 
 Continuous broadband photometry (i.e. VRI) should be made, for a complete
pulsation cycle, about a year.
For both the RCB and AGB stars, a spot computer model will attempt to fit the light variations in various bands to multiple spots on the surface of the star (Strassmeier 1988).  The different points on the surface inside and outside of spots will be assigned different temperatures. The flux is then integrated over the visible hemisphere of the star and then compared to the photometric observations in different bands.  Good fits can be obtained using this method but the results are not unique unless combined with Doppler imaging.
In parallel, a search for magnetic field in, e.g., SiO masers,
should me made.

A protruding cool magnetic spot ($\S 3.3$), when at an angle of 
$\sim 90^\circ$ to the line of sight, will cause the AGB star to
appear as asymmetrical. 
 Speckle interferometry can be used to study such stars
e.g., Karovska {\it et al.} (1991), to detect deviations from symmetry.
Karovska {\it et al.} (1991) mention several possibilities for the
asymmetry they detect in Mira, one of them is a large convective spot
(Schwarzschild 1975). 
 We would like to add to their list a large protruding magnetic spot,
as one of the possibilities of causing deviation from sphericity.

% %=========================================================
\section{SUMMARY}
% %=========================================================

 Our main goals and results can be summarized as follows.
\newline 
(1)  Properties of cool magnetic spots as known from the sun, can naturally
   explain many properties of mass loss from AGB and RCB stars.
   The assumption that we made here is that magnetic dynamo activity
   occurs in these evolved stars even when they rotate very slowly,
   $\sim 10^{-4}$ times their equatorial Keplerian velocity (Soker 1998).
\newline 
(2) We calculate the position of the spots' photosphere. 
    In AGB stars the spots protrude from the photosphere, while they are deeper 
    in the envelope of RCB stars.
\newline 
(3) Using the mechanism proposed by Woitke {\it et al.} (1996), 
    and the results of Frank (1995),
    we suggest that the lower temperature and the magnetic field
    above the spot, facilitate dust formation closer to the 
    stellar surface, after the passage of a shock wave 
    driven by the stellar pulsation.
\newline 
(4) We propose observations that can be made to look for the presence
    of cool magnetic spots on AGB and RCB stars. 
   These include long-term photometric monitoring in several 
   broadband filters, with temporal resolution of days, and 
   speckle interferometry of AGB stars. 
\newline 
(5) Future calculations should combine the work of Frank (1995) and  
    Woitke {\it et al.} (1996) with a magnetic field above the spot.
    We need 2D or, even better, 3D simulations of shock waves propagating 
    from the cool spot and around it, with the magnetic pressure included
    above the cool spot.

{\bf ACKNOWLEDGMENTS:} 
This research was supported in part by a grant from the University of 
Haifa and a grant from the Israel Science Foundation. 
NS would like to thank the Israeli Students Union for their 6 week long strike
during the winter semester, which allowed the completion of this work 
in a relatively short time. GC was supported by NASA grant, JPL 961526.

\end{document}